\documentclass[a4paper,UKenglish]{eptcs}
\usepackage{paralist}

\usepackage{etoolbox} 
\newtoggle{techreport}
\toggletrue{techreport}

\usepackage[utf8]{inputenc}
\usepackage{afterpage}              
\usepackage{amsmath}
\usepackage{amssymb}
\iftoggle{techreport}{
  \usepackage{amsthm}
}{}

\usepackage{bussproofs} 
\EnableBpAbbreviations

\usepackage{cancel}
\usepackage{caption,subcaption}     

\usepackage{cmll} 
\usepackage{colonequals}            
\usepackage{DotArrow} 

\usepackage{enumitem}              

\usepackage{environ}                
\usepackage{etex,etoolbox}          
\usepackage{eucal} 
\usepackage{float} 
\usepackage{framed} 
\usepackage{graphicx}               
\usepackage[protrusion=true,expansion=true]{microtype} 
\usepackage{multirow}               

\usepackage{nicefrac}
\usepackage{pifont} 
\usepackage{proof}                  
\usepackage{relsize} 
\usepackage{rotating}               
\usepackage{stmaryrd}               
\usepackage{keyval}

\usepackage{thmtools, thm-restate, thm-autoref}  
\usepackage{trsym} 
\usepackage[usenames,dvipsnames]{xcolor}

\usepackage{tikz}                   
\usetikzlibrary{matrix, positioning, arrows, shapes}

\usepackage{fixltx2e} 
\usepackage{xspace} 
\usepackage{xifthen} 

\usepackage{wrapfig} 

\usepackage[nomargin,inline,index,draft]{fixme} 
\fxusetheme{color}

\FXRegisterAuthor{fxAS}{anfxAS}{\color{red} {\underline{Alceste}}}
\FXRegisterAuthor{bart}{anbart}{\color{magenta} {\underline{bart}}}
\FXRegisterAuthor{MM}{anMM}{\color{blue} {\underline{Maurizio}}}

\usepackage{mathtools} 

\usepackage{hyperref}
\hypersetup{
  colorlinks   = true, 
  urlcolor     = blue, 
  linkcolor    = blue, 
  citecolor   = red 
}
\hypersetup{final}

\usepackage{cleveref} 

\usepackage{babel} 

\newlist{inlinelist}{enumerate*}{1}
\setlist*[inlinelist,1]{%
  label=(\roman*),
}

\newtheorem{notation}{Notation}
\newtheorem{definition}{Definition}

\newtheorem{proposition}{Proposition}

\newtheorem{example}{Example}
\newtheorem*{nono-definition}{Definition}
\newtheorem*{nono-proposition}{Proposition}

\def\eg{e.g.\@\xspace}%
\def\ie{i.e.\@\xspace}%
\NewEnviron{restheorem}[2][]
  {\begin{restatable}[#1]{theorem}{#2}%
    \label{#2}
    \BODY
   \end{restatable}%
   \iftoggle{techreport}{
     \begin{proof}%
       See 
       page~\pageref{#2-proof}.
     \end{proof}}{}}
\NewEnviron{reslemma}[2][]
  {\begin{restatable}[#1]{lemma}{#2}%
    \label{#2}
    \BODY
   \end{restatable}%
   \iftoggle{techreport}{
     \begin{proof}%
       See 
       page~\pageref{#2-proof}.
     \end{proof}
   }{}}
\NewEnviron{resproposition}[2][]
  {\begin{restatable}[#1]{proposition}{#2}%
    \label{#2}
    \BODY
   \end{restatable}%
   \iftoggle{techreport}{
     \begin{proof}%
       See 
       page~\pageref{#2-proof}.
     \end{proof}
   }{}}

{}

\newenvironment{proofof}[1]{%
  \paragraph{Proof of  \Cref{#1} on \cpageref{#1}}
  \label{#1-proof}
  \csname #1*\endcsname
  \begin{proof}}%
  {\end{proof}}

\newcommand{\suchthat}{\mathbin{.}}%
\newcommand{\setenumSmall}[1]{\mathord{\{#1\}}}%
\newcommand{\setcompSmall}[2]{\mathord{\{{#1} \,\mid\, {#2}\}}}%

\iftoggle{techreport}{%
\newcommand{\setenum}[1]{\mathord{\left\{#1\right\}}}
\newcommand{\setcomp}[2]{\mathord{\left\{{#1} \,\middle|\, {#2}\right\}}}

}{%
\newcommand{\setenum}[1]{\setenumSmall{#1}}
\newcommand{\setcomp}[2]{\setcompSmall{#1}{#2}}

}

\newcommand{\ifempty}[3]{%
  \ifthenelse{\isempty{#1}}{#2}{#3}%
}

\newcommand{\inferrule}[1]{\textsc{\scriptsize (#1)}}
\newcommand{\inference}[3][]{\infer[\ifempty{#1}{}{\inferrule{#1}}]{#3}{#2}}

\newcommand{\relR}{\mathrel{\mathcal{R}}}

\newcommand{\lts}[1]{\mathcal{\uppercase{#1}}}

\def\ltssetColor{OliveGreen}

\newcommand{\BH}{\ensuremath{\mathord{\color{\ltssetColor}{\mathbb{U}}}}}

\newcommand{\Acts}{\mathord{\mathsf{A}}}

\def\ltsLab{\ell}
\def\ltsTau{\tau}
\def\ltsLabTau{\ltsLab_{\ltsTau}}

\newcommand{\Labs}{\Acts} 
\newcommand{\LabsTau}{\Acts_{\ltsTau}} 

\newcommand{\ltsMove}[2][]{\ifempty{#1}{\xrightarrow{#2}}{\mathrel{\xrightarrow{#2}_{\lts{#1}}}}}

\newcommand{\ltsMoveP}[2][]{\mathord{\ltsMove[#1]{#2}}} 
\newcommand{\ltsNotMoveP}[2][]{\mathord{\not\ltsMove[#1]{#2}}} 
\newcommand{\ltsMoveStar}[1]{\mathrel{\ltsMove{}^{\!*}}}
\newcommand{\ltsWMove}[2][]{\ifempty{#1}{\xRightarrow{#2}}{\mathrel{\xRightarrow{#2}_{\lts{#1}}}}}

\newcommand{\ltsWMoveP}[2][]{\mathord{\ltsWMove[#1]{#2}}} 

\newcommand{\ltsBaseDual}[1]{\operatorname{co}({#1})}
\newcommand{\ltsDual}[1]{\ltsBaseDual{#1}}

\newcommand{\ltsPar}{\mathbin{\|}} 
\newcommand{\ltsNil}{\mathbf{0}}

\newcommand{\Success}{\mathcal{S}}

\newcommand{\ltsReadyFmtBase}[4]{\mathord{{#4}{#1}_{#3}^{#2}}}
\newcommand{\ltsReadyFmt}[3]{\ltsReadyFmtBase{\,\downarrow}{#1}{#2}{#3}}
\newcommand{\ltsWReadyFmt}[3]{\ltsReadyFmtBase{\Downarrow}{#1}{#2}{#3}}
\newcommand{\ltsReady}[1]{\ltsReadyFmt{}{}{#1}}
\newcommand{\ltsWReady}[1]{\ltsWReadyFmt{}{}{#1}}
\newcommand{\ltsMayDiverge}[1]{\mathord{{#1}{\,\uparrow}}}

\def\ltsBaseFmt#1{#1}
\newcommand{\ltsSp}[1][]{\ifempty{#1}{\mathord{\ltsBaseFmt{p}}}{\mathord{\ltsBaseFmt{p}_{#1}}}}
\newcommand{\ltsSpi}[1][]{\ifempty{#1}{\mathord{\ltsBaseFmt{p}'}}{\mathord{\ltsBaseFmt{p}'_{#1}}}}
\newcommand{\ltsSq}[1][]{\ifempty{#1}{\mathord{\ltsBaseFmt{q}}}{\mathord{\ltsBaseFmt{q}_{#1}}}}

\newcommand{\ltsSqi}[1][]{\ifempty{#1}{\mathord{\ltsBaseFmt{q'}}}{\mathord{\ltsBaseFmt{q}'_{#1}}}}

\newcommand{\ltsSr}{\mathord{\ltsBaseFmt{r}}}

\newcommand{\atomFmt}[1]{\mathsf{#1}}
\newcommand{\atom}[1]{\atomFmt{#1}}

\newcommand{\chanFmt}[1]{#1}

\def\InTag{\mathord{?}}
\def\OutTag{\mathord{!}}

\newcommand{\atomIn}[1]{\mathord{{\InTag\atom{#1}}}}
\newcommand{\atomOut}[1]{\mathord{{\OutTag\atom{#1}}}}

\newcommand{\complianceF}[1]{\ifempty{#1}{\mathcal{C}}{\mathcal{C}(#1)}}

\newcommand{\compliant}[1][]{\mathrel{\lhd}_{#1}}

\newcommand{\progress}{\compliant[\mathit{pg}]}

\newcommand{\must}{\compliant[\mathit{mst}]}

\newcommand{\should}{\compliant[\mathit{shd}]}

\newcommand{\may}{\compliant[\mathit{may}]}

\newcommand{\iocompliant}{\compliant[\mathit{io}]}

\newcommand{\bhcompliant}{\compliant[\mathit{beh}]}

\newcommand{\chanIOFmt}[3]{\mathord{\chanFmt{#1}\mathbin{\ifempty{#1}{}{\!}{#2}\ifempty{#1}{}{\!}}\atomFmt{#3}}}

\newcommand{\chanIn}[2]{\chanIOFmt{#1}{\InTag}{#2}}
\newcommand{\chanOut}[2]{\chanIOFmt{#1}{\OutTag}{#2}}

\newcommand{\chanWReadyFmt}[3]{\ltsWReadyFmt{#1}{#2}{#3}}

\newcommand{\chanWReadyIn}[2]{\chanWReadyFmt{\InTag}{#2}{#1}}
\newcommand{\chanWReadyOut}[2]{\chanWReadyFmt{\OutTag}{#2}{#1}}

\newcommand{\ActsIn}{\mathord{\Acts{}^{\InTag}}}

\newcommand{\ActsOut}{\mathord{\Acts{}^{\OutTag}}}

\newcommand{\proc}[1]{{\color{\procColor}#1}}

\newcommand{\procMove}[1]{\ltsMove{\proc{#1}}}
\newcommand{\procMoveStar}[1]{\mathrel{\procMove{}^{\!*}}}

\newcounter{beh} 

\togglefalse{techreport}

\title{A Note On Compliance Relations And Fixed Points.\footnote{This work has been partially supported by Aut.\ Reg.\ of Sardinia project \textit{Smart collaborative engineering}. We thank the anonymous reviewers for their useful comments on a previous version of this work.}}
\author{Maurizio Murgia
\institute{Universit\`a degli Studi di Cagliari}
\email{\quad maurizio.murgia@unica.it}}
\def\titlerunning{A note on compliance relations and fixed points.}
\def\authorrunning{Maurizio Murgia}

\begin{document}

\maketitle              

\begin{abstract}
We study compliance relations between behavioural contracts in a syntax independent
setting based on Labelled Transition Systems. We introduce a fix-point based family of 
compliance relations, and show that many compliance relations
appearing in literature belong to this family. 
\end{abstract}

\section{Introduction} \label{sec:intro}
Behavioural contracts are abstract descriptions of the external behaviour and 
interaction scheme of distributed services \cite{HuttelLVCCDMPRT16}. 
They often come together 
with some compliance relation, which intuitively relates contracts
of services whose composition is correct, where the notion of
correctness is specific to the application domain \cite{BartolettiCZ15}.
In a related line of research, so called testing theories are used to study
observational equivalence of CCS processes through the concept of passing a test 
\cite{Nicola84tcs}. Roughly, two processes are equivalent if they pass the same sets of 
tests. Tests are themselves processes, and a process passes a test when its 
parallel composition with the test enjoys some behavioural property 
(\eg, must or may reach a successful state). 
In retrospect, the relation between a process and a passed test can be seen as a
compliance relation \cite{Laneve07concur}. 
A selection of compliance/test relations, and their
relative merits and inclusions, has been surveyed in \cite{BartolettiCZ15} 
in a common ground
based on Labelled Transition Systems. However, there is still lack of a general 
unifying theory of compliance relations, which would help to improve current practices 
in design and implementation of distributed concurrent systems.

\paragraph{Contribution.}
This paper is a first step towards a better understanding of the mathematical 
foundations of compliance relations. The starting point is a simple observation,
based on two well known compliance relations: progress and must compliance. 
Progress relates contracts whose composition never gets stuck, or terminates in a 
successful state. Must relates contracts whose composition always terminates in a 
successful state. Intuitively, there is a duality between progress, which 
allows infinite behaviour, and must, which is only about finite behaviour.
Two standard tools for reasoning about finiteness and infiniteness are, respectively,
induction and coinduction, or, equivalently, least and greatest fixed points of 
monotonic functionals over complete lattices.
This paper introduces a family of compliance relations,
dubbed fix-compliance relations, defined as the set of fixed point 
of a simple and natural functional. We show that progress and must
are, respectively, the greatest and the least fixed point of such compliance 
functional.
We also consider other notions of compliance. For instance, should and behavioural
compliance, which allow for infinite behaviour but with some limitations,
turn out to be intermediate fixed points. Some compliance relations 
in literature are not fix-compliance, e.g. IO-compliance
and may compliance. However, it turn out that IO-compliance is a post-fixed point, 
while may is a pre-fixed point.

\paragraph{Synopsis.}
We start introducing the contract model and some notation in \Cref{sec:contracts}.
We then define the compliance functional and the concept of fix-compliance
in \Cref{sec:coinductive-compliance}. In the rest of 
\Cref{sec:coinductive-compliance} we present several known compliance relations,
and we show how the fit the fix-compliance framework. \Cref{sec:conclusions} 
discusses related works and concludes. Some rappresentative proofs are 
relegated to \Cref{sec:proofs}.


\section{Contracts} \label{sec:contracts}

In this~\namecref{sec:contracts} we present a model of contracts,
following the lines of \cite{BartolettiCZ15}. %
Contracts are formalised as states of a Labelled Transition System (LTS) 
where labels are partitioned into 
\emph{internal}, \emph{input}, and \emph{output} actions. %
All the compliance relations defined later on in~\Cref{sec:coinductive-compliance} 
will be formalised as binary relations between states. %


Our treatment is developed within the LTS %
$\big (\BH, \LabsTau, \setcomp{\ltsMove{\ltsLabTau}}{\ltsLabTau \in \LabsTau} \big)$, %
where: 
\begin{itemize}
\item%
  $\BH$ is the universe of \emph{states} 
  (ranged over by $\ltsSp, \ltsSq, \ldots$), also called \emph{contracts}; %
\item%
  $\LabsTau$ (ranged over by $\ltsLabTau,\ltsLabTau',\ldots$) 
  is the set of \emph{labels}, partitioned into %
  \emph{input actions $\chanIn{}{a}, \chanIn{}{b}, \ldots \in \ActsIn$}, %
  \emph{output actions $\chanOut{}{a}, \chanOut{}{b}, \ldots \in \ActsOut$}, %
  and the \emph{internal action}~$\ltsTau$; %
\item%
  $\ltsMove{\ltsLabTau}\; \subseteq \BH \times \BH$ is 
  a \emph{transition relation}, for all $\ltsLabTau$.
\end{itemize}

We let $\ltsLab,\ltsLab',\ldots$ range over $\Acts = \ActsIn \cup \ActsOut$.
We postulate an involution $\ltsDual{\cdot}$ on~$\Acts$, %
such that $\ltsDual{\chanIn{}{a}} = \chanOut{}{a}$ %
and $\ltsDual{\chanOut{}{a}} = \chanIn{}{a}$. %
The \emph{reducts of $\ltsSp$} are the states reachable from $\ltsSp$ 
with a finite sequence of transitions with any label, while the \emph{$\ltsLabTau$-reducts of $\ltsSp$} 
are the states reachable from $\ltsSp$ with a finite sequence of transitions with label $\ltsLabTau$. %
A \emph{trace} is a (possibly infinite) sequence 
$\ltsSp_0 \ltsMove{{\ltsLabTau}^{\raisebox{-4pt}{\scalebox{0.5}{(1)}}}} \ltsSp_1 \ltsMove{{\ltsLabTau}^{\raisebox{-4pt}{\scalebox{0.5}{(2)}}}} \cdots$. %
A $\ltsTau$-trace  is a trace where
${\ltsLabTau}^{\scalebox{0.6}{$(i)$}} = \ltsTau$, for all $i$
(similarly for $\ltsTau$-reduct). %
We assume that there exists a unique state with no outgoing transitions. Such state is denoted by $\ltsNil$. Note that, since $\ltsNil$ is unique, if $\ltsSp$ is such that $\ltsSp \not\ltsMove{\ltsLabTau}$ for all $\ltsLabTau$, then $\ltsSp = \ltsNil$. 
We interpret $\ltsNil$ as a correctly terminated state, and we will often refer to $\ltsNil$ as the \emph{success} state.
%

\begin{notation} 
  \label{def:outmove}%
  \label{def:inout-proj}
  \label{def:ltsWMove}
  We adopt the following notation: %
  \begin{itemize}
    
  \item $\relR^*$ for the reflexive and transitive closure of a relation $\relR$

  \item $\ltsSp \;\ltsMoveP{\ltsLabTau}\;$ when
    $\exists \ltsSp' \suchthat \ltsSp \ltsMove{\ltsLabTau} \ltsSp'$. %
    Further, we write
    $\ltsSp \;\ltsMoveP{\;\;}\;$ when 
    $\exists \ltsLabTau \suchthat \ltsSp \ltsMoveP{\ltsLabTau}$

  \item for a set $L \subseteq \Labs$, we define
    $L^{\InTag} = L \cap \ActsIn$ and $L^{\OutTag} = L \cap \ActsOut$

  \item $\ltsWMove{} \; = \, (\ltsMove{\ltsTau})^*$ is the \emph{weak transition relation}.
    We define $\ltsWMove{\ltsLabTau}$ as
    \(
    \ltsWMove{} \ltsMove{\ltsLabTau} \ltsWMove{}
    \)

  \item $\ltsReady{\ltsSp} = \setcomp{\ltsLab}{\ltsSp \ltsMoveP{\ltsLab}}$
    are the \emph{barbs of $\ltsSp$},
    and $\ltsWReady{\ltsSp} = \setcomp{\ltsLab}{\ltsSp \ltsWMoveP{\ltsLab}}$
    are its \emph{weak barbs}

  \item $\ltsMayDiverge{\ltsSp}$ is true 
    when $\ltsSp$ has an infinite internal computation
    $\ltsSp \ltsMove{\ltsTau} \ltsSp_1 \ltsMove{\ltsTau} \ltsSp_2 \ltsMove{\ltsTau} \cdots$

  \end{itemize}
  The above notation for $\ltsMoveP{}$ is extended to $\ltsWMoveP{}$ as expected. %
\end{notation}

In order to define parallel composition of contracts, we require some additional structure on $\BH$.
In particular, we assume $\BH$ to be closed under a binary operation $\ltsPar$.
Contracts in the form $\ltsSp \ltsPar \ltsSq$ are called \emph{compositions}, and we refer
to the left component $\ltsSp$ as the \emph{client} and the right component $\ltsSq$ as the \emph{server}.
Compositions where the client is $\ltsNil$ are called \emph{successful} , and we refer to the set of all successful compositions as $\Success$. Formally,
\(
\Success = \setcomp{\ltsNil \ltsPar \ltsSp\;\;}{\;\;\ltsSp \in \BH}
\).
Note that $\ltsNil$ models success of a single participant, while the elements of $\Success$ model
success of compositions of (at least) two participants. Intuitively, $\Success$ contains all compositions in which the client is terminated, and so in which the server has successfully satisfied the client. This asymmetric notion can be found in previous work \cite{Barbanerad15}.

The semantics of compositions formalises the standard synchronisation \`a la CCS~\cite{Milner89ccs}.

\begin{definition}[Parallel composition]
  \label{def:ltspar}%
  For all $\ltsSp, \ltsSq \in \BH$, %
  we impose $\ltsSp \ltsPar \ltsSq \in \BH$. 
  The transition relation of compositions contains all and only the transitions that can be derived
  with the following rules:
  \[
  \inference
  {\ltsSp \ltsMove{\ltsLabTau} \ltsSp'}
  {\ltsSp \ltsPar \ltsSq \ltsMove{\ltsLabTau} \ltsSp' \ltsPar \ltsSq}
  \hspace{30pt}
  \inference
  {\ltsSq \ltsMove{\ltsLabTau} \ltsSq'}
  {\ltsSp \ltsPar \ltsSq \ltsMove{\ltsLabTau} \ltsSp \ltsPar \ltsSq'}
  \hspace{30pt}
  \inference
  {\ltsSp \ltsMove{\ltsLab} \ltsSp' & \ltsSq \ltsMove{\ltsDual{\ltsLab}} \ltsSq'}
  {\ltsSp \ltsPar \ltsSq \ltsMove{\ltsTau} \ltsSp' \ltsPar \ltsSq'}
  \]
\end{definition}







  
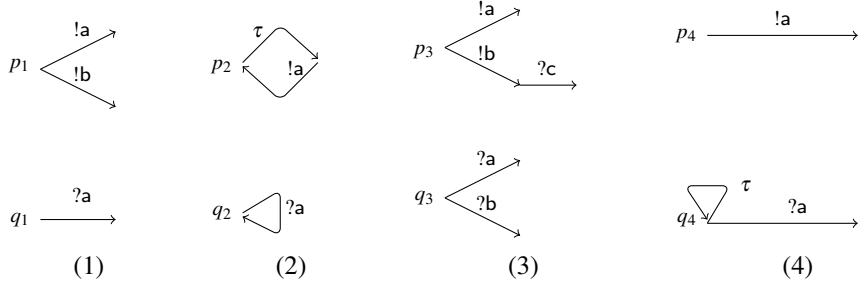
\begin{figure}
  \begin{center}
    \begin{tabular}{c}
      \begin{subfigure}[b]{.15\linewidth}
        \setcounter{subfigure}{\value{beh}}
        \begin{tikzpicture}
          \node [left] at (0,1.05) {\scriptsize$\ltsSp[1]$};

          \draw [->] (0,1) -- (1,1.5);
          \node [above] at (0.55,1.3) {\scriptsize$\atomOut{a}$};
          
          \draw [->] (0,1) -- (1,0.5);
          \node [above] at (0.55,0.7) {\scriptsize$\atomOut{b}$}; 
          
          \node [left] at (0,-1) {\scriptsize$\ltsSq[1]$};
          \draw [->] (0,-1) -- (1,-1);
          \node [above] at (0.55,-0.9) {\scriptsize$\atomIn{a}$};

          
        \end{tikzpicture}

        \caption{} \label{beh:ex1}
      \end{subfigure}

      \hspace{0pt}
      \stepcounter{beh}

      \begin{subfigure}[b]{.15\linewidth}
        \setcounter{subfigure}{\value{beh}}
        \begin{tikzpicture} 
          \node [left] at (0,-0.05) {\scriptsize$\ltsSp[2]$};
          \draw [rounded corners=1mm, ->] (0,0) -- (0.5, 0.5) -- (1,0.05);
          \node [above] at (0.2,0.2) {\scriptsize$\ltsTau$};
          
          \draw [rounded corners=1mm, ->] (1,0) -- (0.5, -0.5) -- (0,-0.05);
          \node [above] at (0.7,-0.3) {\scriptsize$\atomOut{a}$}; 

          \node [left] at (0,-2) {\scriptsize$\ltsSq[2]$};
          \draw [rounded corners=1mm,->] (0,-2) --(0.5, -1.7)--(0.5, -2.3) --  (0,-2.05);
          \node [above] at (0.7,-2.15) {\scriptsize$\atomIn{a}$};

          
        \end{tikzpicture}
        \caption{} \label{beh:ex2} 
      \end{subfigure}
      
      \hspace{0pt}    
      \stepcounter{beh}
      
      \begin{subfigure}[b]{.2\linewidth}
        \setcounter{subfigure}{\value{beh}}
        \begin{tikzpicture}
          \node [left] at (0,-0.05) {\scriptsize$\ltsSp[3]$};
          \draw [->] (0,0) -- (1,0.5);
          \node [above] at (0.55,0.3) {\scriptsize$\atomOut{a}$};
          
          \draw [->] (0,0) -- (1,-0.5);
          \node [above] at (0.55,-0.3) {\scriptsize$\atomOut{b}$}; 
          
          \draw [->] (1,-0.5) -- (1.75,-0.5);
          \node [above] at (1.35,-0.5) {\scriptsize$\atomIn{c}$}; 
          \node [left] at (0,-2) {\scriptsize$\ltsSq[3]$}; 
          \draw [->] (0,-2) -- (1,-1.5);
          \node [above] at (0.55,-1.7) {\scriptsize$\atomIn{a}$};
          \draw [->] (0,-2) -- (1,-2.5);
          \node [above] at (0.55,-2.3) {\scriptsize$\atomIn{b}$};
          
          
        \end{tikzpicture}
        \caption{} \label{beh:may-not-pr:ca}
      \end{subfigure}
      \hspace{0pt}    
      \stepcounter{beh}
     \begin{subfigure}[b]{.22\linewidth}
        \setcounter{subfigure}{\value{beh}}      

        \begin{tikzpicture}
          \node [left] at (0,0) {\scriptsize$\ltsSp[4]$}; 
          \draw [->] (0,0) -- (2,0);
          \node [above] at (1,0) {\scriptsize$\atomOut{a}$};

          \node [left] at (0,-2.45) {\scriptsize$\ltsSq[4]$};
          \draw [->] (0,-2.5) -- (2,-2.5);
          \node [above] at (1.2,-2.5) {\scriptsize$\atomIn{a}$};
          \draw [rounded corners=1mm, ->] (0,-2.5) -- (0.3,-2) -- (-0.3,-2) -- (0,-2.45);
          \node [right] at (0.3,-2) {\scriptsize$\ltsTau$};
          
        \end{tikzpicture}
        \caption{} \label{beh:sh-not-bh:ia}
      \end{subfigure}

    \end{tabular} 
  \end{center}
  \caption{Some pairs of contracts.}
  \label{fig:notcompliant}

\end{figure}

\section{A fixed-point based family of compliance relations}%
\label{sec:coinductive-compliance}





In this section we introduce a general class of compliance relations 
between behaviours,
based on the compliance functional $\complianceF{}$ defined below. We then show that many compliance
relations in literature, but not all, fit within this class. 
Compliance relation in this class have the following properties:
\begin{itemize}
\item contracts whose composition is successful are compliant;
\item compositions of compliant contracts never get stuck before a successful
state is reached;
\item compliance is preserved by $\tau$-transitions, 
until a successful state is reached.
\end{itemize}

\begin{definition}
\label{def:compliance-functional}
We define the compliance functional $\complianceF{}: \BH^2 \rightarrow \BH^2$ as follows:
\[
\complianceF{x} = \Success \cup \setcomp{(\ltsSp,\ltsSq)}{\ltsSp \ltsPar \ltsSq \ltsMoveP{\ltsTau} 
\land (\ltsSp \ltsPar \ltsSq \ltsMoveP{\ltsTau}
\ltsSpi \ltsPar \ltsSqi \implies (\ltsSpi,\ltsSqi) \in x)} 
\] 
We say that a relation $\relR \subseteq \BH^2$ is:
\begin{itemize}
\item a \emph{pre}-compliance relation if $\relR$ is a pre-fixed point of $\complianceF{}$,
that is $\complianceF{\relR} \subseteq \relR$;
\item a \emph{post}-compliance relation if $\relR$ is a post-fixed point of $\complianceF{}$, 
	that is $\relR \subseteq \complianceF{\relR}$;
\item a \emph{fix}-compliance relation if $\relR$ is a fixed-point of $\complianceF{}$,
	that is $\relR = \complianceF{\relR}$.
\end{itemize}
\end{definition}

We start recalling that, by the Knaster-Tarski theorem \cite{tarski1955}, 
every monotonic endo-function over a complete lattice has a least fixed point and a 
greatest fixed point (they may coincide). Furthermore,
the least fixed-point coincides with the least pre-fixed point and the greatest
fixed point coincides with the greatest post-fixed point.
We will now on work on the complete lattice $\BH \times \BH$ ordered by set inclusion.
It is easy to verify that $\complianceF{}$ is monotonic with respect to $\subseteq$, that is,
for all $x,y \subseteq \BH \times \BH$:
\[
	x \subseteq y \implies \complianceF{x} \subseteq \complianceF{y}
\]

\paragraph{Progress compliance.}

We start by considering the notion of \emph{progress},
which consists of absence of deadlocks 
(on the client-side, since we are considering the asymmetric relation). %
Formally, in~\Cref{def:progress} we say that %
a contract $\ltsSp$ has progress with $\ltsSq$ %
(in symbols, $\ltsSp \progress \ltsSq$) %
iff, whenever a $\ltsTau$-reduct of $\ltsSp \ltsPar \ltsSq$ 
is stuck,
then $\ltsSp$ has reached the success state. %

\begin{definition}[Progress]
  \label{def:progress}
  We write $\ltsSp \progress \ltsSq$ iff: %
  \[ %
  \ltsSp \ltsPar \ltsSq 
  \;\ltsWMove{}\; %
  \ltsSp' \ltsPar \ltsSq' 
  \;\ltsNotMoveP{\ltsTau} %
  \;\;\text{ implies }\;\; %
  \ltsSp' = \ltsNil 
  \]
\end{definition}

This notion has been used \eg in $\tau$-less CCS~\cite{Castagna09toplas},
in session types 
(both untimed~\cite{Barbanerad15} and timed~\cite{BartolettiCM17}), %
and in types for CaSPiS~\cite{Acciai08ugo65}. 

\begin{example}
\label{ex:progress}
Consider the behaviours in \cref{fig:notcompliant}.
\begin{itemize}
\item
We have that $\ltsSp[1] \progress \ltsSq[1]$: the composition 
$\ltsSp[1] \ltsPar \ltsSq[1]$ can only $\tau$-reduce 
through a synchronisation on $\atom{a}$, 
leading to a successful state.
\item
The composition $\ltsSp[2] \ltsPar \ltsSq[2]$
can only take the $\ltsSp[2]$ $\tau$-move, and then synchronise
on $\atom{a}$, going back to the starting state. Therefore, 
$\ltsSp[2] \progress \ltsSq[2]$.
\item 
The composition $\ltsSp[3] \ltsPar \ltsSq[3]$ may $\tau$-reduce 
through a synchronisation on $\atom{b}$, 
leading to a state which is stuck (no $\tau$-reductions are possible) but
unsuccessful ($\ltsSp[3]$ is not terminated as she can emit a $\atomIn{c}$ action).
Therefore, $\ltsSp[3] \not\progress \ltsSq[3]$.
\item The composition $\ltsSp[4] \ltsPar \ltsSq[4]$ can loop taking the 
$\ltsSp[4]$ $\tau$-move, or $\tau$-reduce to a successful state
through a synchronisation on $\atom{a}$. Therefore, 
$\ltsSp[4] \progress \ltsSq[4]$.  
\end{itemize}
\end{example}

It turns out that $\progress$ is the largest fix-compliance.

\begin{proposition}\label{th-progress}
$\progress$ is the largest fix-compliance.
\end{proposition}
An important consequence of \Cref{th-progress} is that all post-compliance relations
enjoy the progress property (as defined in \Cref{def:progress}): indeed, if $x$ is a post-compliance, then, by the Knaster-Tarki Theorem it follows $x \subseteq \progress$.
%
%

  \paragraph{Must-testing compliance.}

The notion of compliance in~\cite{Acciai10coordination}
is inspired to must-testing~\cite{Nicola84tcs}. %
Must testing requires 
a contract to reach success in \emph{all} (sufficiently long) traces. %
Formally, we say that a $\ltsTau$-trace 
$\ltsSr_0 \ltsMove{} \ltsSr_1 \ltsMove{} \cdots$ 
is \emph{maximal} if it is infinite, or if it ends in a state
$\ltsSr_n$ such that $\ltsSr_n \ltsNotMoveP{\ltsTau}$. %
A behaviour $\ltsSp$ is must-testing compliant with $\ltsSq$ 
(in symbols, $\ltsSp \must \ltsSq$) if, in all the maximal
$\ltsTau$-traces of $\ltsSp \ltsPar \ltsSq$, 
the contract $\ltsSp$ reaches the $\ltsNil$ state. %

\begin{definition}[Must-testing compliance]
  \label{def:must}
  We write\; $\ltsSp_0 \must \ltsSq_0$ \;iff\; %
  \begin{center}
    for all maximal $\ltsTau$-traces
    \(
    \ltsSp_0 \ltsPar \ltsSq_0
    \ltsMove{\ltsTau} %
    \ltsSp_1 \ltsPar \ltsSq_1
    \ltsMove{\ltsTau} %
    \cdots
    \)
    \; : \;
    \(
    \exists i \geq 0
    \suchthat
    \ltsSp_i  =  \ltsNil
    \)
  \end{center}
\end{definition}

\begin{example}
\label{ex:must}
Consider the behaviours in \cref{fig:notcompliant}.
\begin{itemize}
\item
$\ltsSp[1] \must \ltsSq[1]$: the only maximal $\tau$-trace is
$\ltsSp[1] \ltsPar \ltsSq[1] \ltsMove{\ltsTau} \ltsNil \ltsPar \ltsNil$,
which contains a composition whose left component is $\ltsNil$.
\item
$\ltsSp[2] \not\must \ltsSq[2]$:
the composition $\ltsSp[2] \ltsPar \ltsSq[2]$
diverges without visiting a successful state.
\item 
$\ltsSp[3] \not\must \ltsSq[3]$, basically for the same reason of
\Cref{ex:progress}.
\item $\ltsSp[4] \not\must \ltsSq[4]$:
the composition $\ltsSp[4] \ltsPar \ltsSq[4]$ may perpetually loop taking the 
$\ltsSp[4]$ $\tau$-move, without visiting any successful state.  
\end{itemize} 
\end{example}

\begin{resproposition}{th-must}
$\must$ is the least fix-compliance relation.
\end{resproposition}

  \paragraph{Should-testing compliance.}

We now present a notion of compliance 
inspired by the theory of should-testing~\cite{Brinksma95concur,Rensink07infoco}. %
A behaviour $\ltsSp$ is \emph{should-testing compliant} with $\ltsSq$ 
(in symbols, $\ltsSp \should \ltsSq$) if, 
after every possible \emph{finite} $\tau$-trace of $\ltsSp \ltsPar \ltsSq$, 
there exists a subsequent (finite) $\tau$-trace which leads
$\ltsSp$ to the success state. %

\begin{definition}[Should-testing compliance]
  \label{def:should}
  We write\; $\ltsSp \should \ltsSq$ \;iff\; %
  \[
  \ltsSp \ltsPar \ltsSq 
  \;\ltsWMove{}\; %
  \ltsSp' \ltsPar \ltsSq' %
  \;\;\text{ implies }\;\; %
  \exists \ltsSq''
  \;\suchthat\;
  \ltsSp' \ltsPar \ltsSq' %
  \ltsWMove{}
  \ltsNil \ltsPar \ltsSq''
  \]
\end{definition}

A notion similar to the one in~\Cref{def:should}
has been used in~\cite{Bravetti07fsen} 
(under the name of \emph{correct contract composition}),
and in~\cite{Aalst10cj,BCP15scp} (where it is named \emph{weak termination}).

\begin{example}
\label{ex:should}
Consider the behaviours in \cref{fig:notcompliant}.
\begin{itemize}
\item
$\ltsSp[1] \should \ltsSq[1]$: the composition 
$\ltsSp[1] \ltsPar \ltsSq[1]$ can only $\tau$-reduce 
through a synchronisation on $\atom{a}$, 
leading to a successful state.
\item
$\ltsSp[2] \not\should \ltsSq[2]$.
As noted in \Cref{ex:must}, the composition $\ltsSp[2] \ltsPar \ltsSq[2]$
necessarily diverges, and no successful state is reachable.
\item 
$\ltsSp[3] \not\should \ltsSq[3]$, for the same reason of \Cref{ex:progress,ex:must}.
\item 
$\ltsSp[4] \should \ltsSq[4]$.
The composition $\ltsSp[4] \ltsPar \ltsSq[4]$ can loop taking the 
$\ltsSp[4]$ $\tau$-move, but a successful state is invariantly reachable
through a synchronisation on $\atom{a}$. 
\end{itemize} 
\end{example}
\begin{proposition}
$\should$ is a fix-compliance relation.
\end{proposition}

  \paragraph{Behavioural compliance.}

\Cref{def:bhcompliant} below formalises in our setting 
the relation called \emph{behavioural compliance}
in~\cite{Laneve07concur,Laneve15fac}. %
A contract $\ltsSp$ is compliant with $\ltsSq$ 
(in symbols, $\ltsSp \bhcompliant \ltsSq$), 
if, in every possible $\tau$-reduct
$\ltsSpi \ltsPar \ltsSqi$ of $\ltsSp \ltsPar \ltsSq$, 
two conditions are satisfied:
if the reduct is stuck, then $\ltsSpi$ has reached success;
otherwise,
if $\ltsSqi$ alone can produce an infinite $\ltsTau$-trace,
then $\ltsSpi$ must be able to reach success 
without further synchronisations. %

\begin{definition}[Behavioural compliance]
  \label{def:bhcompliant}
  We write $\ltsSp \bhcompliant \ltsSq$ iff:
  \[
  \ltsSp \ltsPar \ltsSq 
  \;\ltsWMove{}\; 
  \ltsSp' \ltsPar \ltsSq'
  \;\;\text{ implies }\;\; 
  \big(
  \ltsSp' \ltsPar \ltsSq'  \,\ltsNotMoveP{\ltsTau}{}\,
  \text{ implies } 
  \ltsSp' = \ltsNil
  \big)
  \; \land \;
  \big(
  \ltsMayDiverge{\ltsSq'}
  \text{ implies } 
  \ltsSp' \ltsWMove{} \ltsNil
  \big)
  \]
\end{definition}
\begin{example}
\label{ex:bh}
Consider the behaviours in \cref{fig:notcompliant}.
\begin{itemize}
\item
$\ltsSp[1] \bhcompliant \ltsSq[1]$: $\ltsSq[2]$ does not diverge, 
and the composition $\ltsSp[1] \ltsPar \ltsSq[1]$ can only $\tau$-reduce 
through a synchronisation on $\atom{a}$, 
leading to a successful state.
\item
$\ltsSp[2] \bhcompliant \ltsSq[2]$:
as noted in \Cref{ex:progress}, the composition $\ltsSp[2] \ltsPar \ltsSq[2]$
never gets stuck, and $\ltsSq[2]$ does not diverge.
\item 
$\ltsSp[3] \not\bhcompliant \ltsSq[3]$, for the same reason of 
\Cref{ex:progress,ex:must,ex:should}.
\item 
$\ltsSp[4] \not\bhcompliant \ltsSq[4]$:
Although the composition $\ltsSp[4] \ltsPar \ltsSq[4]$ never gets stuck,
$\ltsSq[4]$ may diverge and $\ltsSp[4]$ cannot terminate on her own. 
\end{itemize} 
\end{example}
\begin{proposition}
$\bhcompliant$ is a fix-compliance.
\end{proposition}

  \paragraph{I/O compliance.}

In~\cite{BSZ14concur}, 
a contract $\ltsSp$ is considered compliant with $\ltsSq$ 
(in symbols, $\ltsSp \iocompliant \ltsSq$), 
if, in every possible $\tau$-reduct $\ltsSp' \ltsPar \ltsSq'$ of $\ltsSp \ltsPar \ltsSq$, 
the weak outputs of $\ltsSp'$ are included in the weak inputs of $\ltsSq'$; 
further, if $\ltsSp'$ has no weak outputs but still some weak inputs, 
then they include the weak outputs of~$\ltsSq'$. %

\begin{definition}[I/O compliance]
  \label{def:iocompliant}
  We write $\ltsSp \iocompliant \ltsSq$ iff
  \(
  \ltsSp \ltsPar \ltsSq 
  \ltsWMove{} %
  \ltsSpi \ltsPar \ltsSqi %
  \)
  implies: %
  \begin{align*}
    &
    \chanWReadyOut{\ltsSpi}{} \subseteq \ltsDual{\chanWReadyIn{\ltsSqi}{}} 
    \;\;\land\;\; %
    \big( %
    ( %
    \chanWReadyOut{\ltsSpi}{} = \emptyset %
    \,\land\,
    \chanWReadyIn{\ltsSpi}{} \neq \emptyset %
    ) %
    \;\implies\; %
    \emptyset \neq %
    \chanWReadyOut{\ltsSqi}{} %
    \subseteq 
    \ltsDual{\chanWReadyIn{\ltsSpi}{}} 
    \big) %
  \end{align*}
\end{definition}

\begin{example}
\label{ex:io}
Consider the behaviours in \cref{fig:notcompliant}.
\begin{itemize}
\item
$\ltsSp[1] \not\iocompliant \ltsSq[1]$: $\chanWReadyOut{\ltsSp[1]}{} = 
\setenum{\atomOut{a},\atomOut{b}} \not\subseteq \setenum{\atomOut{a}}= 
\ltsDual{\chanWReadyIn{\ltsSq[1]}{}}$. 
\item
$\ltsSp[2] \iocompliant \ltsSq[2]$:
we have that, in every $\tau$-reduct $\ltsSpi[2] \ltsPar \ltsSqi[2]$
of $\ltsSp[2] \ltsPar \ltsSq[2]$, 
$\chanWReadyOut{\ltsSpi[2]}{} = \setenum{\atomOut{a}}$ and 
$\ltsDual{\chanWReadyIn{\ltsSqi[2]}{}} = \setenum{\atomOut{a}}$.
Therefore both conjuncts of \Cref{def:iocompliant} holds.
\item 
$\ltsSp[3] \not\iocompliant \ltsSq[3]$: after a synchronisation on
$\atom{b}$, a state $\ltsSpi[3] \ltsPar \ltsSqi[3]$ is reached.
However, $\chanWReadyOut{\ltsSpi[3]}{} = \emptyset$ and
$\chanWReadyIn{\ltsSpi[3]}{} \neq \emptyset$, but 
$\chanWReadyOut{\ltsSqi[3]}{} = \emptyset$. Therefore, the second
conjunct of \Cref{def:iocompliant} does not hold.
\item 
$\ltsSp[4] \iocompliant \ltsSq[4]$:
The only reachable states are $\ltsSp[4] \ltsPar \ltsSq[4]$ and 
$\ltsNil \ltsPar \ltsNil$. As 
$\chanWReadyOut{\ltsSpi[2]}{} = \setenum{\atomOut{a}}$ and 
$\ltsDual{\chanWReadyIn{\ltsSqi[2]}{}} = \setenum{\atomOut{a}}$,
$\ltsSp[4] \ltsPar \ltsSq[4]$ satisfies the condition of \Cref{def:iocompliant}.
For $\ltsNil \ltsPar \ltsNil$, it does hold as well:
$\chanWReadyOut{\ltsNil}{} = \emptyset= \ltsDual{\chanWReadyIn{\ltsNil}{}}$.
\end{itemize} 
\end{example}

It turns out that $\iocompliant$ is a post-compliance but not
a pre-compliance (and hence not a fix-compliance). To see why it is not
a pre-compliance, consider $\ltsSp[1]$ and $\ltsSq[1]$ from \cref{fig:notcompliant}.
As noted in \Cref{ex:io}, $\ltsSp[1] \not\iocompliant \ltsSq[1]$. However,
$\ltsSp[1] \ltsPar \ltsSq[1] \ltsMove{\tau}$ and its unique $\tau$-reduct is 
successful and hence composed by compliant behaviours. Therefore, 
$(\ltsSp[1],\ltsSq[1]) \in \complianceF{\iocompliant}$. 

\begin{resproposition}{th-io}
$\iocompliant$ is a post-compliance relation.
\end{resproposition}

  \paragraph{May-testing compliance.}

In~\Cref{def:may}, a contract $\ltsSp$ is said to be
\emph{may-testing compliant} with $\ltsSq$ 
(in symbols, $\ltsSp \may \ltsSq$) 
if there exists a finite $\ltsTau$-trace of $\ltsSp \ltsPar \ltsSq$ 
which leads $\ltsSp$ to the success state. %

\begin{definition}[May-testing compliance]
  \label{def:may}
  We write\; $\ltsSp \may \ltsSq$ \;iff\; %
  \[
  \exists \ltsSq'
  \; \suchthat \;
  \ltsSp \ltsPar \ltsSq 
  \;\ltsWMove{}\; %
  \ltsNil \ltsPar \ltsSq'
  \]
\end{definition}

\begin{example}
\label{ex:may}
Consider the behaviours in \cref{fig:notcompliant}.
\begin{itemize}
\item
$\ltsSp[1] \may \ltsSq[1]$: $\ltsSp[1] \ltsPar \ltsSq[1]$
can reach a successful state after 
a synchronisation on $\atom{a}$.
\item
$\ltsSp[2] \not\may \ltsSq[2]$:
as noted in \Cref{ex:should}, the composition $\ltsSp[2] \ltsPar \ltsSq[2]$
never reach any successful state.
\item 
$\ltsSp[3] \may \ltsSq[3]$: $\ltsSp[3] \ltsPar \ltsSq[3]$
can reach a successful state after 
a synchronisation on $\atom{a}$.
\item 
$\ltsSp[4] \may \ltsSq[4]$:
$\ltsSp[4] \ltsPar \ltsSq[4]$
can reach a successful state after 
a synchronisation on $\atom{a}$.
\end{itemize} 
\end{example}

In a sense, may-testing compliance assumes a \emph{cooperative} scenario:
participants pre-agree on their internal choices,
and the scheduler to only permit the synchronisations leading to success,
seen here as a common goal. %
%

It turns out that $\may$ is a pre-compliance relation but not a post-compliance
relation (and hence not a fix-compliance). To see why it is not a post-compliance,
consider $\ltsSp[3]$ and $\ltsSq[3]$ of \cref{fig:notcompliant}. As noted in
\Cref{ex:may}, $\ltsSp[3] \may \ltsSq[3]$. However, 
$(\ltsSp[3],\ltsSq[3]) \not \in \complianceF{\may}$: through a synchronisation on
$\atom{b}$, $\ltsSp[3] \ltsPar \ltsSq[3]$ can reduce
to a composition which is not successful nor composed by may-compliant behaviours.

\begin{proposition}
$\may$ is a pre-compliance relation.
\end{proposition}

\section{Conclusions and related work.}%
\label{sec:conclusions}
Behavioural contracts and compliance relations have been studied in several works and 
contexts, e.g. service-oriented computing \cite{Acciai08ugo65,Acciai10coordination,
Bravetti07fsen,Castagna09toplas,Laneve07concur,Laneve15fac,Aalst10cj} and session types
\cite{Barbanerad15,BartolettiCM17,BSZ14concur}. Testing preorders have been studied in \cite{Nicola84tcs,Laneve07concur,Rensink07infoco}. The definition of testing compliance in this work is slightly different from the classical ones \cite{Nicola84tcs,Laneve07concur,Rensink07infoco}: 
there, the successful states
are those that can emit the special label $e$. Following \cite{BartolettiCZ15},
we consider $\ltsNil$ as the success state. This makes our treatment simple and uniform. The work \cite{BartolettiCZ15}
presents a taxonomy of compliance relations in a general setting based on LTS similar to the one used in this paper, but they also study certain subclasses of the model,
which correspond to known contract models or process algebras: session types 
\cite{Honda98esop}, $\tau$-less CCS \cite{DeNicola87ccstau}, 
contract automata \cite{Basile14tgc} and interface automata \cite{Alfaro01sigsoft}.
Our work, instead, studies only on the full model, focusing on the mathematical
foundations, and revealing the important role of the compliance functional
$\complianceF{}$. Among the compliance relations surveyed in \cite{BartolettiCZ15},
only IA-compliance (inspired to Interface Automata compatibility) does not seem to be related to $\complianceF{}$ in any way. This seems due to the fact that Interface Automata, being naturally suited for modelling systems composed of many components, do not fit well our binary setting. 

We have introduced a family of compliance relations, 
showing how different treatments to divergences in distributed systems
correspond to different fixed-point of a general functional.
In particular:
\begin{itemize}
\item
Must compliance, which disallows any form of divergence, is
the least fixed-point of $\complianceF{}$.
\item
Should compliance relates contracts whose composition may diverge,
but only if a successful terminated state is always reachable. In a sense,
should assumes fairness (but not full cooperation) of participants and the scheduler 
to reach a success state.
This form of fairness is captured as an intermediate fixed-point of $\complianceF{}$.
\item
Behavioural compliance relates contracts whose composition may diverge,
but forbids situations in which divergence of the server disallows
the client to successfully terminate. In this case the server is 
considered adversarial. Also this compliance
is an intermediate fixed-point of $\complianceF{}$.
\item
Progress compliance allows any form of divergence, and is indeed
the greatest fixed point of $\complianceF{}$.  
\end{itemize} 
We have shown two examples of compliance relations appearing in literature that
are not fixed-point of $\complianceF{}$, but turn out to be pre- or post-fixed point of it.
Post-compliance relations, like IO-compliance, still guarantee the good behavioural 
properties reported in \Cref{sec:coinductive-compliance}, namely stuck-freedom and
preservation of compliance by $\tau$-reduction, but somehow relate fewer contracts than
expected. In the specific case of IO-compliance, this is caused by the asymmetric 
treatment of outputs and inputs. The case of may compliance,
is quite enigmatic: may compliance, being ``cooperative'' in nature 
\cite{BartolettiCZ15}, is out of the scope of fix-compliance relations, 
which are biased towards the 
non-cooperative scenario, but still may is a pre-compliance, and so fits somehow in our 
setting. It is still unclear to us whether this can lead to
useful consequences, or it holds just by coincidence.

A possible future direction is the study of cooperative compliance relations
through fixed-points. For instance, we would expect may compliance to be the least
fixed-point of some suitable functional. We expect the greatest fixed-point of such 
functional to be a kind of cooperative progress, relating contracts whose 
composition produces at least one execution which is infinite or terminates in a 
successful state.
An interesting but challenging future direction is characterising the subcontract 
preorders \cite{Castagna09toplas} induced by fix-compliance relations. 

\bibliography{main}
\newpage
\begin{appendix}
%
%

\section{Proofs}
\label{sec:proofs}

%

\begin{proofof}{th-must}
According to the Knaster-Tarski theorem, it suffice to show that $\must$ is
the least pre-fixed point of $\complianceF{}$. In turn, this can proved by showing that
must is a pre-fixed point of $\complianceF{}$, and that any other pre-fixed point of $\complianceF{}$ 
is larger than $\must$.
\begin{itemize}
\item
For the first part, we have to show that $\complianceF{\must} \subseteq \must$.
So, let $(\ltsSp,\ltsSq) \in \complianceF{\must}$. 
If $(\ltsSp,\ltsSq) \in \Success$, it must be $\ltsSp = \ltsNil$, and so for every maximal
$\tau$-trace $\ltsSp[0] \ltsPar \ltsSq[0] \ltsMove{\tau} \ltsSp[1] \ltsPar \ltsSq[1]
\ltsMove{\tau} \hdots$, with $\ltsSp[0] = \ltsSp$ and $\ltsSq[0] = \ltsSq$,
there is $i$ such that $\ltsSp[i] = \ltsNil$: just take $i = 0$.
If $(\ltsSp,\ltsSq) \not\in \Success$, by definition of $\complianceF{}$, we have that 
$\ltsSp \ltsPar \ltsSq
\ltsMove{\tau}$, and, for all $\ltsSpi \ltsPar \ltsSqi$ such that $\ltsSp \ltsPar \ltsSq
\ltsMove{\tau} \ltsSpi \ltsPar \ltsSqi$, it holds that $\ltsSpi \must \ltsSqi$ (1).
Let $\ltsSp \ltsPar \ltsSq \ltsMove{\tau} \ltsSp[1] \ltsPar \ltsSq[1]
\ltsMove{\tau} \hdots$ be a maximal trace. Note that, by (1), it holds that 
$\ltsSp[1] \must \ltsSq[1]$. Therefore, the maximal trace $\ltsSp[1] \ltsPar \ltsSq[1]
\ltsMove{\tau} \hdots$ eventually reaches a state whose left contract is
$\ltsNil$, and thus also the trace $\ltsSp \ltsPar \ltsSq \ltsMove{\tau} \ltsSp[1] 
\ltsPar \ltsSq[1] \ltsMove{\tau} \hdots$, as required.
\item For the second part, let $X$ be a pre-fixed point of $\complianceF{}$, \ie
$\complianceF{X} \subseteq X$. We have to show $\must \subseteq X$.
So, let $\ltsSp \must \ltsSq$. If $\ltsSp = \ltsNil$, it must be 
$(\ltsSp,\ltsSq) \in \Success$,
and so, by definition of $\complianceF{}$, it follows $(\ltsSp,\ltsSq) \in \complianceF{X}$. Since
$\complianceF{X} \subseteq X$ by assumption, we have that $(\ltsSp,\ltsSq) \in X$, as required.
If $\ltsSp \neq \ltsNil$, first note that it must be 
$\ltsSp \ltsPar \ltsSq \ltsMove{\tau}$. Indeed, if this is not the case, the only
maximal $\tau$-trace starting from $\ltsSp \ltsPar \ltsSq$ would be 
$\ltsSp \ltsPar \ltsSq$ (seen as a singleton trace), which of course does not reach
a success state as $\ltsSp$ is not $\ltsNil$ by assumption. Now suppose, by 
contradiction, $(\ltsSp,\ltsSq) \not\in X$. Note that, since
$\complianceF{X} \subseteq X$, it must be $(\ltsSp,\ltsSq) \not\in \complianceF{X}$. Therefore,
by definition of $\complianceF{}$, there must be $(\ltsSp[1],\ltsSq[1]) \not\in X$ such that
$\ltsSp \ltsPar \ltsSq \ltsMove{\tau} \ltsSp[1] \ltsPar \ltsSq[1]$. As
before, $(\ltsSp[1],\ltsSq[1]) \not\in \complianceF{X}$. Iterating the argument again and again, 
we can construct an infinite maximal $\tau$-trace $\ltsSp[0] \ltsPar \ltsSq[0] 
\ltsMove{\tau} \ltsSp[1] \ltsPar \ltsSq[1] \ltsMove{\tau} \hdots$ 
(with $\ltsSp[0] = \ltsSp$ and $\ltsSq[0] = \ltsSq$) such that 
$(\ltsSp[i],\ltsSq[i]) \not\in \complianceF{X}$ for all $i$. Then, since $\Success \subseteq \complianceF{X}$ 
by definition of $\complianceF{}$, it must be $(\ltsSp[i],\ltsSq[i]) \not\in \Success$ for all $i$.
Then, there is maximal $\tau$-trace starting from $\ltsSp \ltsPar \ltsSq$ that
do not reach a success state, contradicting the hypothesis that $\ltsSp \must \ltsSq$.
\end{itemize}
\end{proofof}

\begin{proofof}{th-io}
We have to show that $\iocompliant \subseteq \complianceF{\iocompliant}$.
So, let $\ltsSp \iocompliant \ltsSq$. The case where $\ltsSp = \ltsNil$ is immediate,
as $(\ltsSp,\ltsSq) \in \Success \subseteq \complianceF{\iocompliant}$. For the remaining case 
$\ltsSp \neq \ltsNil$, we have to show that $\ltsSp \ltsPar \ltsSq \ltsMove{\tau}$
and that for all $\ltsSpi,\ltsSqi$ such that $\ltsSp \ltsPar \ltsSq \ltsMove{\tau}
\ltsSpi \ltsPar \ltsSqi$ it holds that $\ltsSpi \iocompliant \ltsSqi$. 
Since $\ltsSp \neq \ltsNil$, it must be 
$\ltsSp \ltsMove{\ltsLabTau}$ for some $\ltsLabTau$. So, if $\ltsLabTau = \tau$,
the thesis follows by the first rule of parallel composition.
If $\ltsLabTau \in \ActsOut$, then $\ltsLabTau \in \chanWReadyOut{\ltsSp}{}$
and so, by the first conjunct in the definition of $\iocompliant$, we have that 
$\chanWReadyOut{\ltsSp}{} \subseteq \ltsDual{\chanWReadyIn{\ltsSq}{}}$. 
Therefore, $\ltsSq \ltsWMoveP{\ltsDual{\ltsLabTau}}$.
If the first transition of such reduction is $\ltsTau$, the thesis follows by an 
application of the second rule of parallel composition. If the first transition
is $\ltsLabTau$, the thesis follows by an application of the third rule.
In the remaining case $\ltsLabTau \in \ActsIn$, we have that 
$\ltsLabTau \in \chanWReadyIn{\ltsSp}{} \neq \emptyset$. So, if 
$\chanWReadyOut{\ltsSp}{} \neq \emptyset$, we can conclude similarly to the previous 
case. If $\chanWReadyOut{\ltsSp}{} = \emptyset$, by the second conjunct in the definition
of $\iocompliant$, we can conclude that 
$\ltsDual{\ltsLabTau} \in \chanWReadyIn{\ltsSp}{}$.
We can then conclude similarly to the previous case.
It remain to show that for all $\ltsSpi,\ltsSqi$ such that $\ltsSp \ltsPar \ltsSq 
\ltsMove{\tau} \ltsSpi \ltsPar \ltsSqi$ it holds $\ltsSpi \iocompliant \ltsSqi$. 
But this follows immediately by the definition of $\iocompliant$,
because $\ltsSpi \ltsPar \ltsSqi$ is a $\tau$-reduct of $\ltsSp \ltsPar \ltsSq$.
\end{proofof}

\end{appendix}

\end{document}
